\documentclass[aps,prl,superscriptaddress,twocolumn]{revtex4-2}

\usepackage{graphicx}
\usepackage{amsmath}
\usepackage{amssymb}
\usepackage{hyperref}
\usepackage[utf8]{inputenc}
\usepackage{mathtools}
\usepackage[english]{babel}
\usepackage{bbm}
\hypersetup{colorlinks=true, linkcolor=blue, citecolor=blue, urlcolor=blue}
\usepackage{xcolor}
\usepackage{braket}
\usepackage[normalem]{ulem}

\newcommand{\ie}{i.\,e.,\ }

\newcommand{\re}{\mathrm{Re}}

\newcommand{\Tr}{\operatorname{Tr}}

\newcommand{\fref}[1]{\text{Fig.}~\ref{#1}}

\newcommand{\eref}[1]{\text{Eq.}~\eqref{#1}}

\begin{document}

\title{Dynamic population of multiexcitation subradiant states in incoherently excited atomic arrays
}

\author{Oriol Rubies-Bigorda}
\email{orubies@mit.edu}
\thanks{equal contributor}
\affiliation{Physics Department, Massachusetts Institute of Technology, Cambridge, Massachusetts 02139, USA}
\affiliation{Department of Physics, Harvard University, Cambridge, Massachusetts 02138, USA}
\author{Stefan Ostermann}
\email{stefanostermann@g.harvard.edu}
\thanks{equal contributor}
\affiliation{Department of Physics, Harvard University, Cambridge, Massachusetts 02138, USA}
\author{Susanne F. Yelin}
\affiliation{Department of Physics, Harvard University, Cambridge, Massachusetts 02138, USA}

\begin{abstract}
The deterministic generation of multiexcitation subradiant states proves to be challenging. Here, we present a viable path towards their transient generation in finite-sized ordered arrays of dipole-dipole coupled quantum emitters, based on incoherent driving of the atomic ensemble. In particular, we show that a maximal coupling to long-lived subradiant states is achieved if only half of the atoms are initially excited. We characterize the nature of the resulting states by calculating the dynamic fluorescence spectrum of the emitted light. Finally, we elucidate the role of coherent interactions during the decay process of sufficiently dense atomic arrays, which result in a coherently driven radiation burst that leads to a subsequent reduction of the chances to prepare multiexcitation subradiant states.
\end{abstract}

\maketitle

\begin{figure}[t]
    \centering
    \includegraphics[width = 0.98\columnwidth]{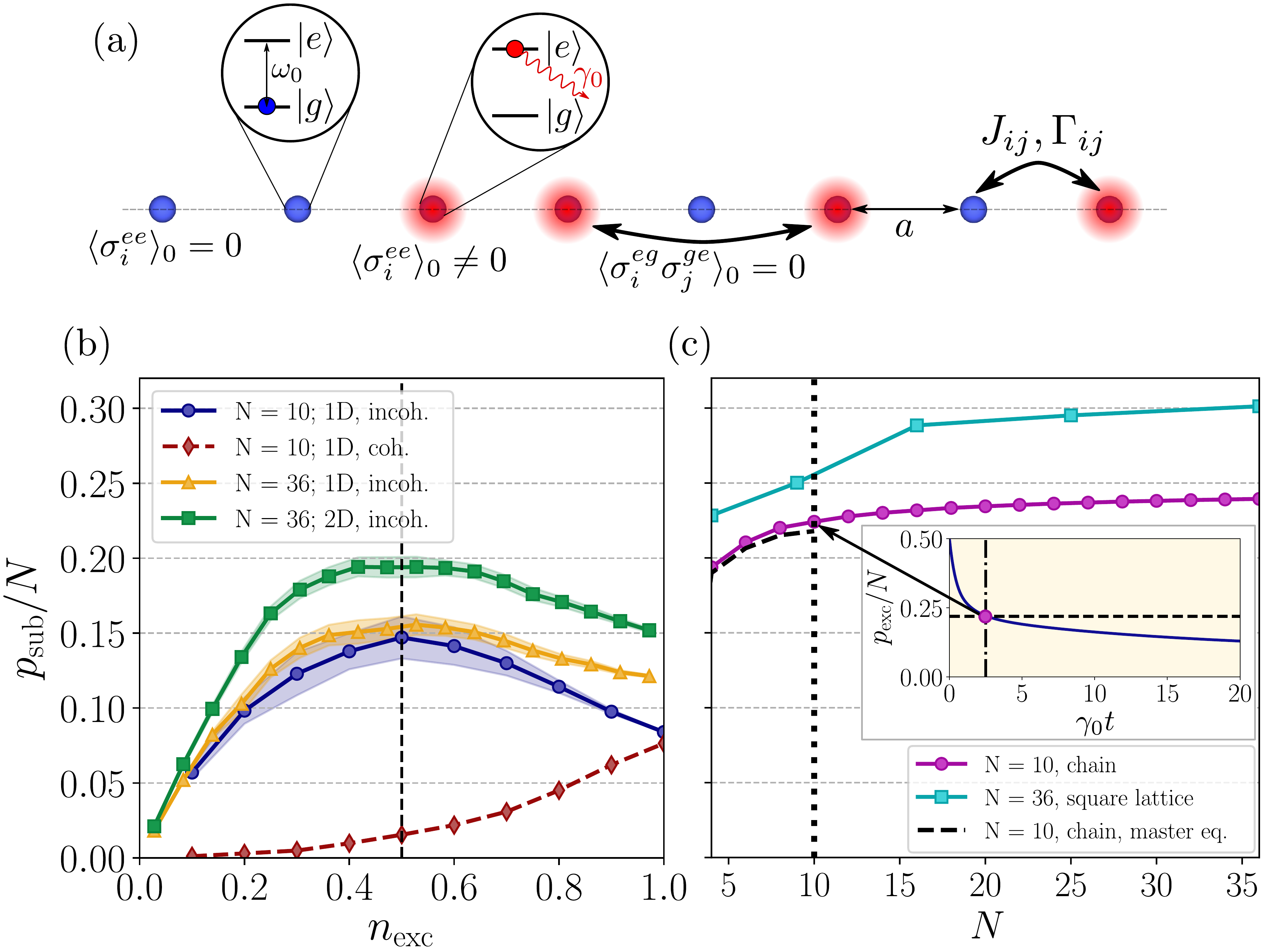}
    \caption{(a) Sketch of a periodic chain of atoms with no pair-correlations and only certain atoms excited (marked in red) at $t=0$. (b) Magnitude of the subradiant population per particle $p_{\mathrm{sub}}/N$,~\ie the excited population left in the array by the time the instantaneous decay rate $\gamma_\mathrm{inst} = 0.1$, as a function of excitation density $n_\mathrm{exc}$ for different array sizes and geometries. Shaded regions encompass one standard deviation. The dashed and solid lines correspond to coherent excitation and incoherent excitation, respectively. (c) $p_{\mathrm{sub}}/N$ as a function of atom number $N$ for a one-dimensional chain (purple) and a two-dimensional square lattice (cyan) for an initial checkerboard excitation distribution. Third order cumulant expansions exhibit good agreement with the master equation solutions (black dashed line).
    The inset shows the decay of the atomic population for a one-dimensional chain of ten atoms, as well as the time at which the system is considered to become subradiant ($\gamma_\mathrm{inst}  =  0.1$) and $p_\mathrm{sub}$ is extracted. The lattice spacing is $a  = 0.15\lambda_0$ in panels (b) and (c).}
    \label{fig:fig1}
\end{figure}
\emph{Introduction}. Recent developments in controlling and manipulating atomic ensembles in predefined geometries~\cite{barredo_an_2016,endres_atom_2016,barredo_synthetic_2018} open up promising avenues towards well-controlled cooperative interactions between light and matter, which are expected to be a fundamental building block for future quantum technologies~\cite{hammerer_quantum_2010}.

If the density of the atomic ensemble is increased such that the spatial separation between atoms is smaller than or on the order of the atomic transition wavelength, light-induced dipole-dipole interactions~\cite{Lehmberg_1970_1,Lehmberg_1970_2} give rise to intriguing cooperative effects such as super- and subradiance~\cite{dicke_coherence_1954,gross_superradiance_1982,bienaime_controlled_2012,guerin_subradiance_2016,footnote_2}. State-of-the-art experiments~\cite{olmos_long-range_2013,rui_subradiant_2020,zhang_2021_2021} are now able to reach this high-density regime, which has triggered numerous studies of subradiance in ordered ensembles of quantum emitters, as well as of its applications in quantum metrology and sensing, quantum information processing and efficient photon storage~\cite{facchinetti_storing_2016, asenjo-garcia_exponential_2017,hebenstreit_subradiance_2017,guimond_subradiant_2019,zhang_theory_2019,Masson_superradiance_nature,rubies-bigorda_photon_2022,ballantine_quantum_2021,ferioli_storage_2021,pineiro_orioli_emergent_2022,zanner_coherent_2022,reitz_cooperative_2022}. While most of these works focus on the single-excitation manifold where only one individual photon or excitation is shared among all atoms at a time, the preparation and analysis of subradiant states with multiple excitations has proven elusive over the years.
This is due to the unfavorable scaling of the Hilbert space, as well as to the complex and non-integrable nature of the underlying spin model, which make a thorough study of large ensembles of emitters difficult.

Here, we study this challenging multiexcitation regime and focus on the transient generation of multiexcitation subradiant states in periodic atomic arrays in free space.
The combination of the Monte Carlo wave function (MCWF) method~\cite{dum_monte_1992,molmer_monte_1993} to solve the master equation and a cumulant expansion of the Heisenberg-Langevin equations up to third order~\cite{kubo_generalized_1962, plankensteiner2022quantumcumulants} allows us to analyze large particle numbers in the multiexcitation regime, as well as to characterize and understand the mechanisms leading to many-body subradiance.
Direct addressing of individual subradiant states requires local phase and amplitude imprinting via the driving field at each atomic position. While this is feasible for small numbers of qubits coupled to waveguides, where driving can occur both through the waveguide and through external sideports~\cite{zanner_coherent_2022}, it turns out to be elusive in free-space setups. Therefore, alternative paths towards the dynamic population of subradiant states have to be determined.

In this Letter, we outline two fundamental criteria that need to be fulfilled to dynamically populate multiexcitation subradiant states without requiring single site addressability. First, the initially prepared state has to have a large overlap with the least radiative states of the Lindbladian spectrum~\footnote{This overlap is clearly always maximal for individual subradiant states contained in the spectrum. However, addressing these states is elusive in free-space platforms, where incoming fields only predominantly couple to the superradiant modes of the atomic ensemble.}. We show that this can be achieved by exiting half of the atoms with no initial coherences among them, which does not require single site addressability~\cite{supplement}. In this case, the system naturally evolves into a mixture of multiexcitation subradiant states. The second criterium is that the distance between atoms must be large enough such that the interaction-induced energy shifts do not lead to a population transfer from subradiant to superradiant states. If the second condition is not fulfilled, the dynamic population of bright states due to the coherent part of the atomic interactions gives rise to a rapid build-up of the atomic coherences, and the subsequent appearance of a coherently-driven superradiant burst. This results in an optimal geometry or lattice spacing for which the population of multiexcitation subradiant states is maximized.

\emph{Model}. We consider an ensemble of $N$ identical two-level atoms with resonance frequency $\omega_0 = 2 \pi c / \lambda_0$ that interact with the three-dimensional vacuum radiation field. 
Tracing out the photonic degrees of freedom under the Born-Markov approximation, one obtains the master equation for the atomic density matrix $\hat{\rho}$ \cite{Lehmberg_1970_1,Lehmberg_1970_2, CohenTanoudgi_book}
\begin{equation}
\label{eq:EOM_densitymatrix}
    \frac{d\hat{\rho}}{dt} = -\frac{i}{\hbar} \left[\hat{\mathcal{H}} , \hat{\rho}\right] + \mathcal{L}[\hat{\rho}],
\end{equation}
where the Hamiltonian $\hat{\mathcal{H}}$ describes the coherent interactions between emitters
\begin{equation}
\label{eq:Hamiltonian}
    \hat{\mathcal{H}} = \hbar \omega_0 \sum_{n=1}^N \hat{\sigma}_n^{ee} + \hbar \sum_{n,m \neq n}^N J_{nm}  \hat{\sigma}_n^{eg} \hat{\sigma}_m^{ge}, 
\end{equation}
and the Lindbladian $\mathcal{L}[\hat{\rho}]$ characterizes the dissipative interactions
\begin{equation}
\label{eq: Linbladian}
    \mathcal{L}[\hat{\rho}]  = \! \sum_{n,m=1}^N \! \frac{\Gamma_{nm}}{2} \left( 2 \hat{\sigma}_n^{ge} \hat{\rho} \hat{\sigma}_m^{eg} - \hat{\sigma}_n^{eg} \hat{\sigma}_m^{ge} \hat{\rho} - \hat{\rho} \hat{\sigma}_n^{eg} \hat{\sigma}_m^{ge} \right).
\end{equation}
Here, $\hat{\sigma}_m^{eg} = |e_m \rangle \langle g_m |$ ($\hat{\sigma}_m^{ge} = |g_m \rangle \langle e_m |$) is the raising (lowering) operator for atom $m$, and the coherent and dissipative parts of the dipole-dipole interactions mediated by the vacuum electromagnetic field are given by $J_{nm} - i \Gamma_{nm}/2 = -\frac{3\pi \gamma_0}{\omega_0} {\mathbf{d}^\dagger} \textbf{G}(\textbf{r}_{nm}, \omega_0) \mathbf{d},$~\cite{Lehmberg_1970_1,Lehmberg_1970_2}, where $\mathbf{d}$ is the transition dipole moment of the atoms, $\mathbf{G} (\mathbf{r},\omega)$ is the Green's tensor for a point dipole in vacuum~\cite{Chew_dyadicGreens,dyadic_novotny_hecht_2006,supplement}, and $\mathbf{r}_{nm}=\mathbf{r}_n - \mathbf{r}_m$ is the vector connecting atoms $n$ and $m$. $\Gamma_{nn}=\gamma_0$ is the spontaneous decay rate of a single atom. The Lamb shift $J_{nn}$ is included in the definition of the transition frequency $\omega_0$.

Typically, the dynamics of decaying atomic ensembles are characterized by the excited-state population $p_\mathrm{exc}(t)= \Tr \{\hat{\rho}(t) \sum_n \hat{\sigma}_n^{ee}\} = \sum_n \langle \hat{\sigma}_n^{ee} \rangle (t)$ and the total photon emission rate $\gamma_\mathrm{tot} = - \dot{p}_\mathrm{exc}$. Here, we introduce the instantaneous decay rate, 
\begin{equation}
\label{eq:gamma_inst}
    \gamma_\mathrm{inst} \equiv \frac{\gamma_\mathrm{tot}}{p_\mathrm{exc}} = \gamma_0 + \frac{\sum_{n,m \neq n} \Gamma_{nm} \langle \hat{\sigma}_n^{eg} \hat{\sigma}_m^{ge} \rangle}{\sum_n \langle \hat{\sigma}_n^{ee} \rangle},
\end{equation}
as the figure of merit for characterizing photon emission. 
Unlike $\gamma_\mathrm{tot}$, $\gamma_\mathrm{inst} (t)$ is constant for a pure exponential decay and directly reflects the superradiant ($\gamma_\mathrm{inst} > \gamma_0$) or subradiant ($\gamma_\mathrm{inst} < \gamma_0)$ character of the state $\hat{\rho}(t)$ at each instant. In particular, the deviation from independent decay ($\gamma_\mathrm{inst}=\gamma_0$) is determined by the second term in~\eref{eq:gamma_inst} and arises from the buildup of two-body coherences $\langle \hat{\sigma}_n^{eg} \hat{\sigma}_m^{ge} \rangle$.

We employ two different numerical methods to compute the dynamics of the atomic system. For small system sizes containing up to 10 atoms, we use the MCWF technique to obtain the atomic density matrix governed by the master equation~\eqref{eq:EOM_densitymatrix}~\cite{dum_monte_1992,molmer_monte_1993}. In addition, we perform a cumulant expansion of the Heisenberg-Langevin equations up to third order. To this end, we derive the equations of motion for the expectation values $\langle \dot{\hat{O}} \rangle = \Tr \{ \dot{\hat{\rho}} \hat{O} \}$ of all operators $\hat{O}$ containing at most three atomic operators, i.e., $\langle \hat{\sigma}_i^{ee} \hat{\sigma}_j^{eg} \hat{\sigma}_k^{ge} \rangle$, and expand the averages of fourth-order operator products in terms of products of third-, second-, and first-order expectation values~\cite{kubo_generalized_1962, kramer_generalized_2015, Robicheaux_cumulants,plankensteiner2022quantumcumulants,our_PRA_superradiance,supplement, Cumulant_Kubo}. 
This approximate method allows us to study systems containing up to 36 atoms with remarkable accuracy. This is a three times larger system size than what can be simulated using the MCWF method.

\emph{Generating subradiant states}. To characterize the many-body nature of the dynamically generated subradiant states, we define the subradiant population $p_\mathrm{sub}$ as the total excited-state population left in the array at the time $t_\mathrm{sub}$ at which the instantaneous decay rate reaches $\gamma_\mathrm{inst}=0.1\gamma_0$. As illustrated in the inset of~\fref{fig:fig1}(c), this marks the point in time at which the decay of the excited-state population has drastically slowed down, indicating subradiance.

We calculate this subradiant population as a function of the excitation density $n_\mathrm{exc}\coloneqq N_\mathrm{exc}/N$, where $N$ is the total number of atoms and $N_\mathrm{exc}$ denotes the number of initially excited emitters. First, we choose a coherent spin state of the form $\ket{\psi_\mathrm{coh}} = \prod_n \left( \sqrt{1-n_\mathrm{exc}} \ket{g_n} + e^{i \mathbf{k} \mathbf{r}_n} \sqrt{n_\mathrm{exc}} \ket{e_n} \right)$ as an initial state, which can be experimentally prepared by a coherent laser pulse impinging on the atomic array~\cite{Masson_superradiance_nature} and is typically used to study subradiance in atomic gases~\cite{ferioli_storage_2021, cipris_subradiance_2021}.
For any value of $n_\mathrm{exc}$, the initial excited-state population is coherently shared among the atoms. While we choose $\mathbf{k}=\mathbf{0}$ for the remainder of this work, the presented results generally hold for all $\mathbf{k}$ within the light cone---defined as $|k|<2\pi/\lambda_0$--- that is, for any value of $\mathbf{k}$ that can be achieved experimentally.
The red dashed line in~\fref{fig:fig1}(b) shows the subradiant population obtained using the MCWF approach for a chain of ten atoms prepared in $\ket{\psi_\mathrm{coh}}$.
In this case, the subradiant population is maximal for a fully inverted system ($n_\mathrm{exc} = 1.0$). This phenomenon can be understood by noting that such a coherent initial state predominantly overlaps with the radiative states of the Dicke ladder~\cite{supplement}. Decreasing $n_{\mathrm{exc}}$ simply reduces the overlap with highly excited radiative states, which consequently diminishes the chance that the excitation gets trapped in subradiant states while cascading down the ladder.
Note also that, for coherent initial states $\ket{\psi_\mathrm{coh}}$, the maximum subradiant population is well below $10\%$ of the total atom number. That is, once the system becomes subradiant, there is on average less than one excitation left in the system.

The value of $p_\mathrm{sub}$ can be increased if an optimized initial state is used. In particular, we find that this is the case for partially and incoherently excited arrays,~\ie for initial states of the form  $|\psi_\mathrm{incoh} \rangle = \prod_{n \in \mathcal{E}} \hat{\sigma}_n^{eg} |G \rangle$, where $|G \rangle$ corresponds to the state where all atoms are in the ground state and $\mathcal{E}$ denotes the set of initially excited atoms. Unlike $|\psi_\mathrm{coh}\rangle$, $|\psi_\mathrm{incoh}\rangle$ has no correlations at initial times and only atomic populations are nonzero at $t\!=\!0$ \cite{supplement}. This state can thus be experimentally realized by either destroying the spatial coherence of the impinging laser via a speckle pattern or by applying a large detuning on random atoms during the coherent excitation pulse (see Supplemental Material~\cite{supplement}).

The solid lines in~\fref{fig:fig1}(b) show the subradiant population for different lattice dimensions and sizes averaged over fifty random distributions of incoherent excitations.
$p_\mathrm{sub}$ is maximal if half the atoms are incoherently excited ($n_\mathrm{exc}=0.5$), both for one-dimensional chains and two-dimensional square lattices, and larger values are obtained in the case of two-dimensional geometries. 
In any case, the fraction of atoms that remain excited for long times is on the order of $15$--$20\%$ of the total atom number, which is substantially higher than for coherently excited arrays [red dashed lines in~\fref{fig:fig1}(b)].

The improved behavior of the incoherent initial condition can be understood from the spectrum of the Lindbladian. As opposed to $\ket{\psi_\mathrm{coh}}$, $\ket{\psi_\mathrm{incoh}}$ has an overlap with all states within its corresponding excitation manifold, which ultimately increases the probability of dynamically populating subradiant states. The fact that the maximum $p_\mathrm{sub}$ is reached at $n_\mathrm{exc} = 0.5$ also makes sense intuitively, as it corresponds to the excitation manifold with the largest number of states. As a result, the overlap of the initial state with the most radiative decay channels is minimal and the probability to dynamically reach subradiant states maximal.

One can further increase the subradiant population by determining an initial state that has a larger overlap with the least radiative decay channels than the random configurations considered in~\fref{fig:fig1}(b). Based on intuition gained from the single-excitation manifold, where the most subradiant state is always a checkerboard pattern of positive and negative phases, we now choose initial states where only atoms located at even lattice sites are excited initially. This state can still be readily prepared without single-site addressability (see Supplemental Material for details~\cite{supplement}). In~\fref{fig:fig1}(c), we show $p_{\mathrm{sub}}$ as a function of atom number $N$. The achieved subradiant population is substantially larger than the maxima observed in~\fref{fig:fig1}(b), and reaches values well above $20\%$ of the total atom number even for small systems. That is, at the time $t_\mathrm{sub}$ where the instantaneous decay rate is below $0.1 \gamma_0$, a chain with ten atoms has an average of more than two excitations left. This illustrates the efficient population of two-excitation subradiant states.

\begin{figure}
    \centering
    \includegraphics[width = \columnwidth]{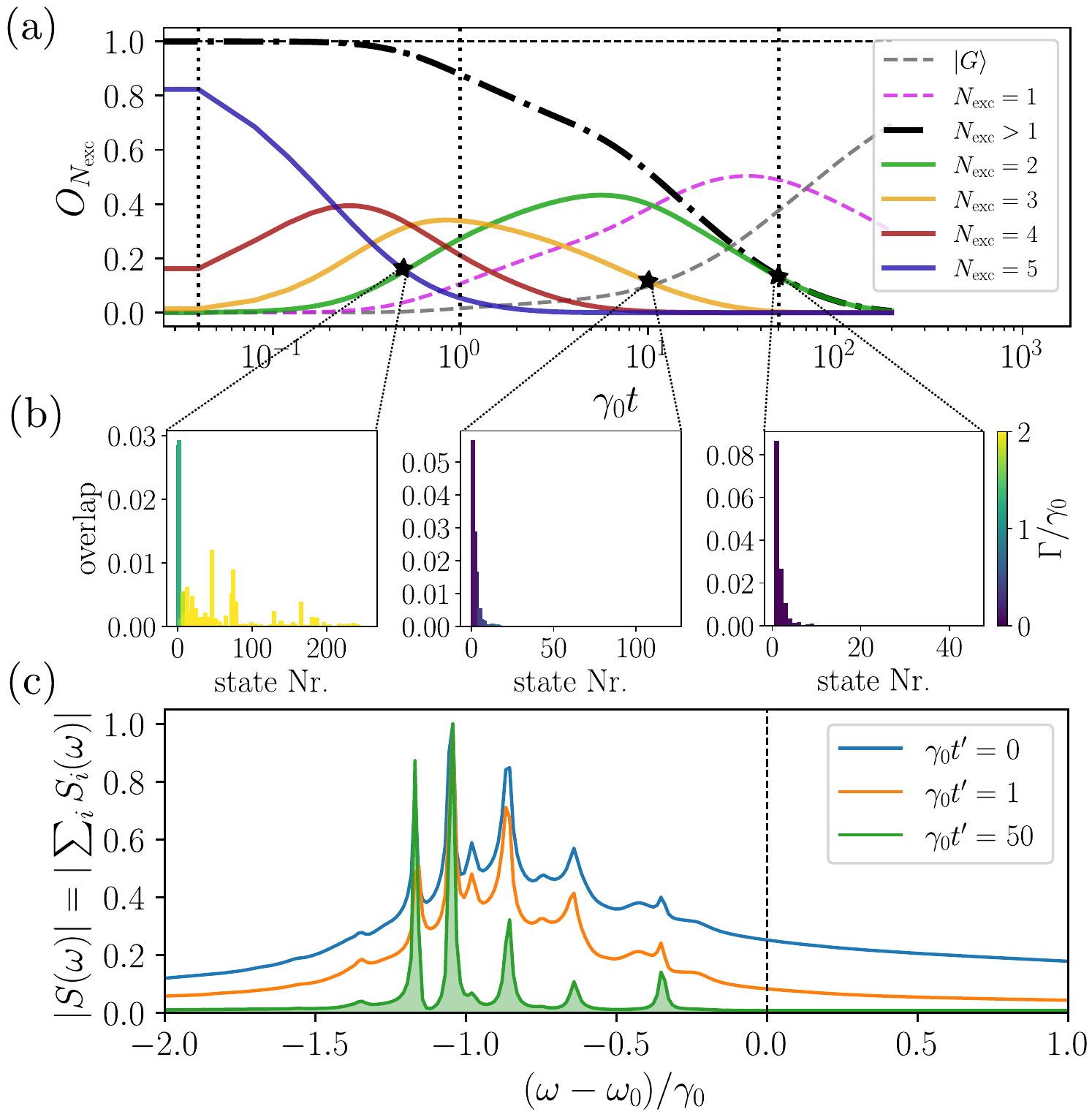}
    \caption{(a) Overlap with the different excitation manifolds over time for a ten-atom chain ($a= 0.15\lambda_0$), with half the atoms initially excited in a checkerboard pattern. A significant contribution from multiexcitation subradiant states ($N_\mathrm{exc}  > 1$) is observed at times $t' \gg 1/\gamma_0$. (b) Overlap of the dynamically populated state at the times indicated by black stars in panel (a) with each individual state contained in a given excitation manifold. Only the most subradiant states are dynamically populated. (b) Instantaneous emission spectrum at different times of the decay process. At early times (blue curve), the fast decaying superradiant states generate a broad background. The most subradiant modes persist at late times (green solid curve) and result in a discrete set of very narrow lines at fixed frequencies.}
    \label{fig:overlaps}
\end{figure}
\emph{Spectrum}. To quantify the population that is dynamically trapped in each excitation manifold, we additionally compute the overlap $O_{N_\mathrm{exc}}(t)=\sum_{\ket{\psi_i}\in\Psi_{N_\mathrm{exc}}}\bra{\psi_i}\hat{\rho}(t)\ket{\psi_i}$ of the state $\hat{\rho}(t)$ with the set of eigenstates of the Hamiltonian containing $N_\mathrm{exc}$ excitations, $\Psi_{N_\mathrm{exc}} = \{\ket{\psi_1}...\ket{\psi_M}\}$. For an atomic array initially prepared in an incoherent checkerboard configuration, the overlap of the many-body state with manifolds containing more than one excitation ($N_\mathrm{exc}>1$) is finite at long times, as shown by the black dashed-dotted curve in~\fref{fig:overlaps}(a). This indicates that the system naturally evolves into a mixture of multiexcitation subradiant states, even for moderate array sizes of just ten atoms. As shown in~\fref{fig:overlaps}(b) and in the supplemental material~\cite{supplement}, the dynamically generated states exhibit a large overlap with the most subradiant states in each excitation manifold.
This effect is particularly pronounced due to the employed checkerboard initial state, as can be seen by comparing our results to a recent work studying subradiant state generation using a statistical mixture as an initial state~\cite{cipris_subradiance_2021, santos_generating_2022}. 

A relevant experimental observable that characterizes the subradiant nature of the state $\hat{\rho}$ is the \emph{dynamic} fluorescence spectrum $S(\omega,t')$. If measured along the direction perpendicular to the array, the spectrum is simply given by the Fourier transform of the two-time correlation function,~\ie $S(\omega,t') = \sum_i S_n(\omega,t') = \sum_n \! 2\re \! \left[\int_{0}^\infty \! d\tau e^{-i\omega\tau}\langle\sigma_n^{eg}(t' \! \! + \! \tau)\sigma_n^{ge}(t') \rangle\right]$~\cite{glauber_coherent_1963}. In~\fref{fig:overlaps}(b), we plot the dynamic spectrum for different times $t'$ at which the spectrum measurement begins. At early times (blue and orange curves), the fast-decaying superradiant states result in a broad background. The narrow peaks correspond to the long-lived subradiant states that are dynamically populated during the decay process. The late-time spectrum obtained at a finite time $t' \gg 1/\gamma_0$ does not contain any contribution from the initial superradiant decay. Hence, the broad background gets strongly suppressed and only the narrow lines remain in the spectrum. Interestingly, the frequencies of these lines do not change over time and are simply determined by the energy shifts associated to the populated subradiant eigenstates of the Hamiltonian.
They can therefore be employed in cooperatively enhanced sensing protocols.

\begin{figure}
    \centering
    \includegraphics[width = 0.95 \columnwidth]{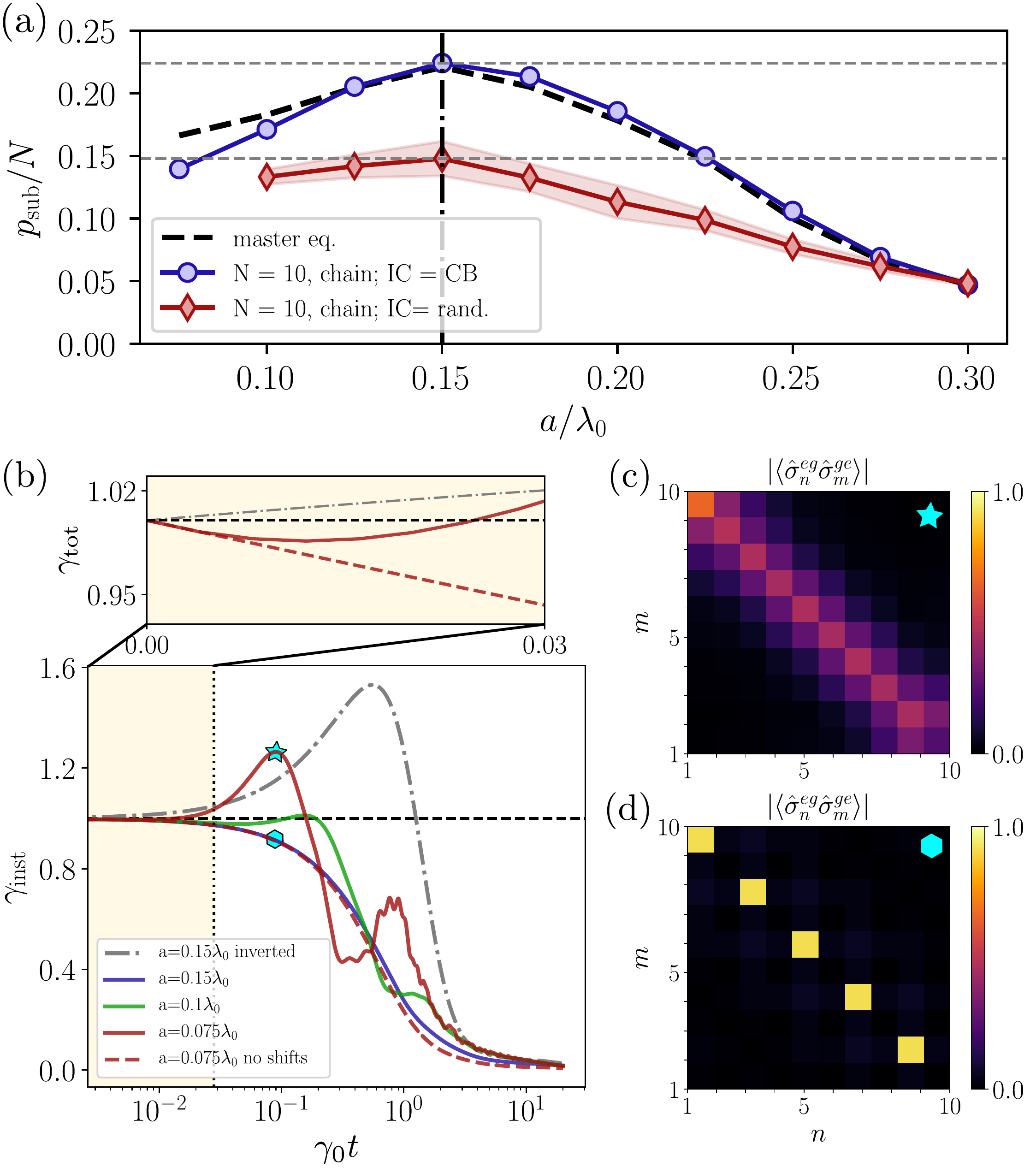}
    \caption{(a) Subradiant population $p_{\mathrm{sub}}$ for an atomic chain with ten atoms as a function of lattice spacing $a$ for a checkerboard exciation pattern (blue) and sets of five randomly excited atoms (red).
    The black dashed line is obtained via the master equation. (b) Normalized emission rate as a function of time for atomic chains with ten atoms and a checkerboard excitation pattern. A radiation burst emerges for small lattice spacings $a = 0.075 \lambda_0$ (red curve), and disappears if coherent dipole-dipole shifts are turned off (dashed red curve). No peak appears for larger spacings ($a = 0.1 \lambda_0$ in solid green and $a = 0.15 \lambda_0$ in solid blue). For comparison, we also plot the superradiant peak for a fully inverted array with $a = 0.15 \lambda_0$ (dashed-dotted grey curve). The upper panel shows the total photon emission rate $\gamma_\mathrm{tot}$ at early times. (c)-(d) Populations (diagonal values) and two-body coherences $\langle \hat{\sigma}_n^{eg} \hat{\sigma}_m^{ge} \rangle$ at the time of the burst for an atomic chain with $a = 0.075 \lambda_0$ (c) with ($J_{ij} \neq 0$) and (d) without ($J_{ij}=0$) coherent shifts.}
    \label{fig: coherent_superradiance}
\end{figure}

\emph{Role of coherent dynamics}. Cooperative effects typically become stronger for decreasing lattice constant $a$. In particular, the Lindbladian in~\eref{eq: Linbladian} approaches the Dicke limit for $a / \lambda_0 \! \rightarrow \! 0$.
Intuitively, this suggests that the overlap of the initial state $\ket{\psi_\mathrm{incoh}}$ with the subradiant manifold increases for decreasing $a$ and that the subradiant population $p_{\mathrm{sub}}$ consequently increases, as shown in~\fref{fig: coherent_superradiance}(a) for atomic chains with $a > 0.15 \lambda_0$. If one further decreases the lattice spacing, however, the coherent dipole-dipole interactions in~\eref{eq:Hamiltonian} become the largest energy scale of the system and start inducing a strong coupling between different states in the same excitation manifold. This results in a population transfer from subradiant to superradiant states, which ultimately reduces $p_{\mathrm{sub}}$ for small $a$ [see~\fref{fig: coherent_superradiance}(a)]. Thus, there is not only an optimal initial condition to dynamically populate subradiant states, but also an optimal lattice spacing or geometry.

Additionally, the appearance of large coherent interactions modifies the emission properties at early times. As shown in the top panel of~\fref{fig: coherent_superradiance}(b), the total emission rate $\gamma_\mathrm{tot}$ initially decreases for states with $n_\mathrm{exc} = N/2$ independent of lattice spacing. That is, the dissipative channels of the system cannot generate a fast build-up of atomic coherences to trigger the onset of a radiation burst. While this results in a monotonic decrease of the total emission at early times for $a > 0.15 \lambda_0$, a radiation peak emerges for small enough lattice spacing [see the solid red curve in~\fref{fig: coherent_superradiance}(b)]. This radiation burst originates from an excitation transfer from subradiant to superradiant states, mediated by the coherent interactions between atoms. The burst vanishes if coherent interactions ($J_{nm}=0$) are artificially put to zero, as evinced by the dashed red curve in~\fref{fig: coherent_superradiance}(b).

The effect can also be understood based on the two-body correlation matrix $\langle \hat{\sigma}_n^{eg} \hat{\sigma}_m^{ge} \rangle$ at the instant where the burst takes place. As shown in~\fref{fig: coherent_superradiance}(c) and (d), the coherences required to observe a peak are only dynamically generated in the presence of coherent dipole-dipole interactions ($J_{nm}\neq 0$). This phenomenon is therefore different in nature from standard Dicke superradiance, where the build-up of correlations and the appearance of a radiation peak occurs due to collective dissipation, and we hereby refer to it as ``coherently-driven superradiance". Additionally, these findings show that the two-photon correlation function at zero time~\cite{Masson_superradiance_nature} can fail to capture the existence of radiation bursts for certain initial conditions of the atomic array. Finally, it is worth noting that the Hamiltonian evolution of the system can be partially engineered by adding spatial modulations of the atomic detunings, which modify the coupling between dark and bright states~\cite{ballantine_quantum_2021,rubies-bigorda_photon_2022} and can consequently enhance or suppress the coherently driven superradiant peak~\cite{supplement}.

\emph{Conclusions and Outlook}. We introduced a viable path towards the dynamic population of multiexcitation subradiant states in atomic emitter arrays that does not require local phase imprinting~\cite{zanner_coherent_2022}. Our approach is based on determining an experimentally feasible initial configuration such that the dynamics results in multiexcitation subradiance. We show that incoherently exciting half of the atoms in the array and choosing a sufficiently large lattice spacing lead to a significant subradiant state population at late times. The resulting states can be characterized by means of the dynamic fluorescence spectrum, which features a peak with a narrow linewidth for each subradiant state populated at late times. These states with cooperatively reduced linewidths are a promising resource for future quantum sensing protocols involving subwavelength emitter arrays. To obtain a good estimate of achievable sensitivities, a detailed study of realistic coherence times, the stability with respect to position fluctuations of the atoms, and the role of lattice vacancies is required~\cite{degen_quantum_2017}.

We further show that a smaller atom spacing does not necessarily lead to improved multiexcitation subradiance. This occurs due to an increase of the coherent dipole-dipole interactions, which strongly couple subradiant and superradiant states and may result in a coherently driven superradiant outburst. Due to its coherent nature, this effect can be compensated and engineered by applying local ac Stark shifts to different atoms in the array \cite{Superlattices_1,Superlattices_2}. Combining such atomic detuning patterns with tailored driving fields remains an important, mostly unexplored avenue to prepare multiexcitation subradiant states~\cite{Stannigel_2012_drivensubradiance,subradiance_driven_waveguide}.

\begin{acknowledgments}
\emph{Acknowledgments}. We would like to thank Valentin Walther and Yidan Wang for fruitful discussions.
O.R.B. acknowledges support from the CUA, as well as from Fundación Mauricio y Carlota Botton and from Fundació Bancaria “la Caixa” (LCF/BQ/AA18/11680093). S.O. is supported by a postdoctoral fellowship of the Max Planck Harvard Research Center for Quantum Optics. SFY would like to acknowledge funding from NSF through the PHY-2207972 and the QSense QLCI as well as from AFOSR. 

O.R.B. and S.O. contributed equally to this work.
\end{acknowledgments}


\bibliographystyle{apsrev4-1-title}
\bibliography{reference_PRL_subradiance}

\begin{thebibliography}{48}
\bibitem [{foo()}]{footnote_2}%
  \BibitemOpen
  \href@noop {} {\bibinfo  {journal} {Temporal superradiance can also occur for
  distant non-interacting atoms provided that the required quantum correlations
  are generated by conditional photon measurements
  \cite{temporal_super_noninter}}\ }\BibitemShut {NoStop}%
\end{thebibliography}

\pagebreak

\clearpage
\onecolumngrid
\begin{center}

\newcommand{\beginsupplement}{%
        \setcounter{table}{0}
        \renewcommand{\thetable}{S\arabic{table}}%
        \setcounter{figure}{0}
        \renewcommand{\thefigure}{S\arabic{figure}}%
     }
\textbf{\large Supplemental Material}
\end{center}
\newcommand{\beginsupplement}{%
        \setcounter{table}{0}
        \renewcommand{\thetable}{S\arabic{table}}%
        \setcounter{figure}{0}
        \renewcommand{\thefigure}{S\arabic{figure}}%
     }
\setcounter{equation}{0}
\setcounter{figure}{0}
\setcounter{table}{0}   
\setcounter{page}{1}
\makeatletter
\renewcommand{\theequation}{S\arabic{equation}}
\renewcommand{\thefigure}{S\arabic{figure}}
\renewcommand{\bibnumfmt}[1]{[S#1]}
\renewcommand{\citenumfont}[1]{S#1}
\vspace{0.8 in}
\newcommand{\D}{\Delta}
\newcommand{\tD}{\tilde{\Delta}}
\newcommand{\K}{K_{PP}}
\newcommand{\bn}{\bar{n}_P}
\newcommand{\G}{\Gamma}
\newcommand{\LH}{\underset{L}{H}}
\newcommand{\HL}{\underset{H}{L}}
\vspace{-1in}

\section{Green's tensor}
\label{SM: Greens}

The Green's tensor for a point dipole in free space which determines the coherent ($J_{nm}$) and dissipative parts ($\Gamma_{nm}$) of the dipole-dipole interactions can be written in Cartesian coordinates as \cite{Lehmberg_1970_1,Lehmberg_1970_2}

\begin{equation}
G_{\alpha \beta}(\mathbf{r},\omega) = \frac{e^{i k r}}{4\pi r} \left[ \left( 1 + \frac{i}{kr} - \frac{1}{(kr)^2} \right) \delta_{\alpha\beta} \right.  + \left. \left(-1 - \frac{3i}{kr} + \frac{3}{(kr)^2} \right) \frac{r_\alpha r_\beta}{r^2} \right] + \frac{\delta_{\alpha \beta} \delta^{(3)}(\mathbf{r})}{3k^2},
\end{equation}

\noindent where $k=\omega/c$, $r=|\mathbf{r}|$, and $\alpha,\beta=x,y,z$.

\section{Third order cumulant expansion}

A system of $N$ two-level atoms has a Hilbert space of size $2^N$. The rapidly growing size of the density matrix ---or wave function, in the case Monte Carlo wave function techniques are used--- required to describe it limits the maximum system size that can be numerically studied to about sixteen atoms. In order to study larger systems, we hereby use cumulant expansions, a method based on neglecting higher order quantum correlations to reduce the number of variables needed to describe the system. 

Using the equation of motion for the density matrix of the system, given by Eq.\,(1) in the main text, one can derive the differential equations describing the dynamics of the expectation values of all products of system operators up to a certain order

\begin{equation}
    \langle \dot{\hat{O}} \rangle = \Tr \{ \dot{\hat{\rho}} \hat{O} \}.
\end{equation}

We consider initial states with no coherences between atoms, which can simply be described as a set of raising operators $\hat{\sigma}_m^{eg} = |e_m \rangle \langle g_m |$ applied to the total ground state of the system $|G \rangle$, that is, $|\psi_\mathrm{incoh} \rangle = \prod_{m \in E} \hat{\sigma}_m^{eg} |G \rangle$. Here, $E$ denotes the set of initially excited atoms. Then, the only non-zero expectation values of the initial state up to third order are 

\begin{subequations}
\label{eq: initial_conditions_2nd_order_cumulants}
\begin{align}
    \langle \hat{\sigma}_m^{ee} \rangle (t=0) &= \left \{
    \begin{aligned}
        &1, && \text{if}\ m \in E  \\
        &0, && \text{otherwise}
    \end{aligned} \right. \\
     \langle \hat{\sigma}_n^{ee} \hat{\sigma}_m^{ee} \rangle (t=0) &= \left \{
    \begin{aligned}
        &1, && \text{if atom}\ i,j \in E\ \\
        &0, && \text{otherwise}
    \end{aligned} \right. \\
     \langle \hat{\sigma}_i^{ee} \hat{\sigma}_j^{ee} \hat{\sigma}_k^{ee}  \rangle (t=0) &= \left \{
    \begin{aligned}
        &1, && \text{if atom}\ i,j,k \in E \\
        &0, && \text{otherwise,}
    \end{aligned} \right. 
\end{align} 
\end{subequations}

\noindent where $\hat{\sigma}_m^{ee} = |e_m \rangle \langle e_m |$. One can further show that the only additional expectation values that become non-zero during the time-evolution of the system are $\langle \hat{\sigma}_i^{eg} \hat{\sigma}_j^{ge} \rangle$ and $\langle \hat{\sigma}_i^{ee} \hat{\sigma}_j^{eg} \hat{\sigma}_k^{ge} \rangle$ \cite{our_PRA_superradiance}, where $\hat{\sigma}_m^{ge} = |g_m \rangle \langle e_m |$ corresponds to the lowering operator of atom $m$.

These expectation values are coupled to fourth-order products of atomic operators. The cumulant expansion consists on approximating such four-atom expectation values as a function of three-, two- and one-atom expectation values as follows \cite{plankensteiner2022quantumcumulants, Cumulant_Kubo}

\begin{align}
\langle \hat{O}_1 \hat{O}_2 \hat{O}_3 \hat{O}_4 \rangle &= \langle \hat{O}_1 \rangle \langle \hat{O}_2 \hat{O}_3 \hat{O}_4 \rangle + \langle \hat{O}_2 \rangle \langle \hat{O}_1 \hat{O}_3 \hat{O}_4 \rangle + \langle \hat{O}_3 \rangle \langle \hat{O}_1 \hat{O}_2 \hat{O}_4 \rangle + \langle \hat{O}_4 \rangle \langle \hat{O}_1 \hat{O}_2 \hat{O}_3 \rangle \nonumber \\
&+ \langle \hat{O}_1 \hat{O}_2 \rangle \langle \hat{O}_3 \hat{O}_4 \rangle + \langle \hat{O}_1 \hat{O}_3 \rangle \langle \hat{O}_2 \hat{O}_4 \rangle + \langle \hat{O}_1 \hat{O}_4 \rangle \langle \hat{O}_2 \hat{O}_3 \rangle \nonumber \\
&-2 \langle \hat{O}_1 \rangle \langle \hat{O}_2 \rangle \langle \hat{O}_3 \hat{O}_4 \rangle -2 \langle \hat{O}_1 \rangle \langle \hat{O}_3 \rangle \langle \hat{O}_2 \hat{O}_4 \rangle -2 \langle \hat{O}_1 \rangle \langle \hat{O}_4 \rangle \langle \hat{O}_2 \hat{O}_3 \rangle -2 \langle \hat{O}_2 \rangle \langle \hat{O}_3 \rangle \langle \hat{O}_1 \hat{O}_4 \rangle \nonumber \\
&-2 \langle \hat{O}_2 \rangle \langle \hat{O}_4 \rangle \langle \hat{O}_1 \hat{O}_3 \rangle -2 \langle \hat{O}_3 \rangle \langle \hat{O}_4 \rangle \langle \hat{O}_1 \hat{O}_2 \rangle + 6 \langle \hat{O}_1 \rangle \langle \hat{O}_2 \rangle \langle \hat{O}_3 \rangle \langle \hat{O}_4 \rangle.  
\end{align}

\noindent One then obtains a closed system of approximately $N^3$ equations, which can be numerically solved for larger atom numbers than with the master equation or the Monte Carlo wave function method \cite{our_PRA_superradiance,Robicheaux_cumulants}.

\section{Additional overlap data}
\begin{figure}
    \centering
    \includegraphics[width = 0.85\columnwidth]{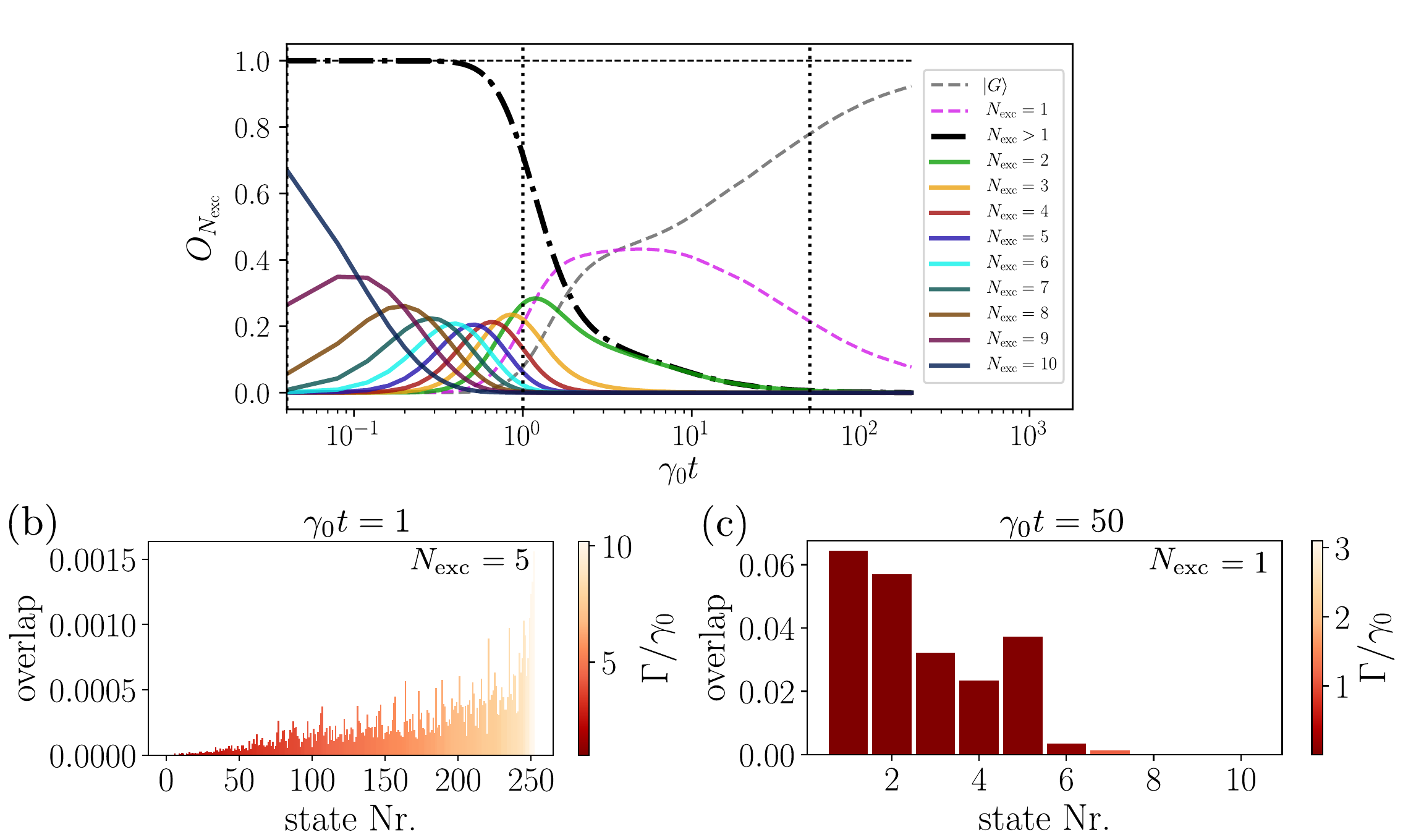}
    \caption{(a) Overlap with the different excitation manifolds over time for a fully inverted ten-atom chain, which corresponds to the optimal configuration to achieve subradiance for a coherent initial excitation [see red dashed line in Fig.\,1(b)]. The dynamically trapped population in multi-excitation subradiant states ($N_\mathrm{exc}>1$) at times larger than $1/\gamma_0$ is significantly smaller than for the incoherent intial excitation in Fig.\,2(a). (b) Overlap of $\hat{\rho} $ at $t \gamma_0=1$ with each individual state of the $N_\mathrm{exc} = 5$ manifold. (c) Overlap with individual single excitation subradiant states ($N_\mathrm{exc} = 1$) at late times $t\gamma_0=50$. The color coding of the bars in panels (b) and (c) displays the collective decay rate each individual state. The states are sorted with increasing collective decay rate $\Gamma$ from left to right. All parameters are the same as in Fig.\,2 of the main text.}
    \label{fig:coherent_overlaps}
\end{figure}

In Fig.\,2(a) of the main text, we consider a ten-atom chain with half the atoms initially excited in a checkerboard pattern $|\psi_\mathrm{incoh} \rangle = \prod_{i} \hat{\sigma}_{2i}^{eg} |G \rangle$ and show its overlap with each excitation manifold $\Psi_{N_\mathrm{exc}}$ as a function of time. For comparison, we provide the data for a coherently excited ensemble of the form $\ket{\psi_\mathrm{coh}} = \prod_n \left( \sqrt{1-n_\mathrm{exc}} \ket{g_n} + e^{i \mathbf{k} \mathbf{r}_n} \sqrt{n_\mathrm{exc}} \ket{e_n} \right)$ in \fref{fig:coherent_overlaps}(a). In this case, the fully inverted array ($n_\mathrm{exc}=1$) results in the largest subradiant population $p_\mathrm{sub}$, as demonstrated in Fig.\,1(b). Still, multiply-excited states ($N_\mathrm{exc} > 1$) are only populated at much shorter times than in the case of the incoherent checkerboard excitation presented in the main text. This occurs because the dynamics of coherently excited arrays are dominated by the most radiative decay channels until the system reaches the single excitation manifold, as evinced by the much faster cascade to lower excitation manifolds in \fref{fig:coherent_overlaps}(a) as opposed to Fig.\,2(a). This trend is confirmed by the overlap of $\hat{\rho}$ with each individual state $\ket{\psi_i}$ making up the higher excitation manifolds $\Psi_\mathrm{N_\mathrm{exc}>1}$. For example, $\ket{\psi_\mathrm{coh}}$ mostly overlaps with the most radiative states in the fifth-excitation manifold at $t=1/\gamma_0$ [see \fref{fig:coherent_overlaps}(b)] resulting in fast radiative decay.

\begin{figure}
    \centering
    \includegraphics[width = 0.7\columnwidth]{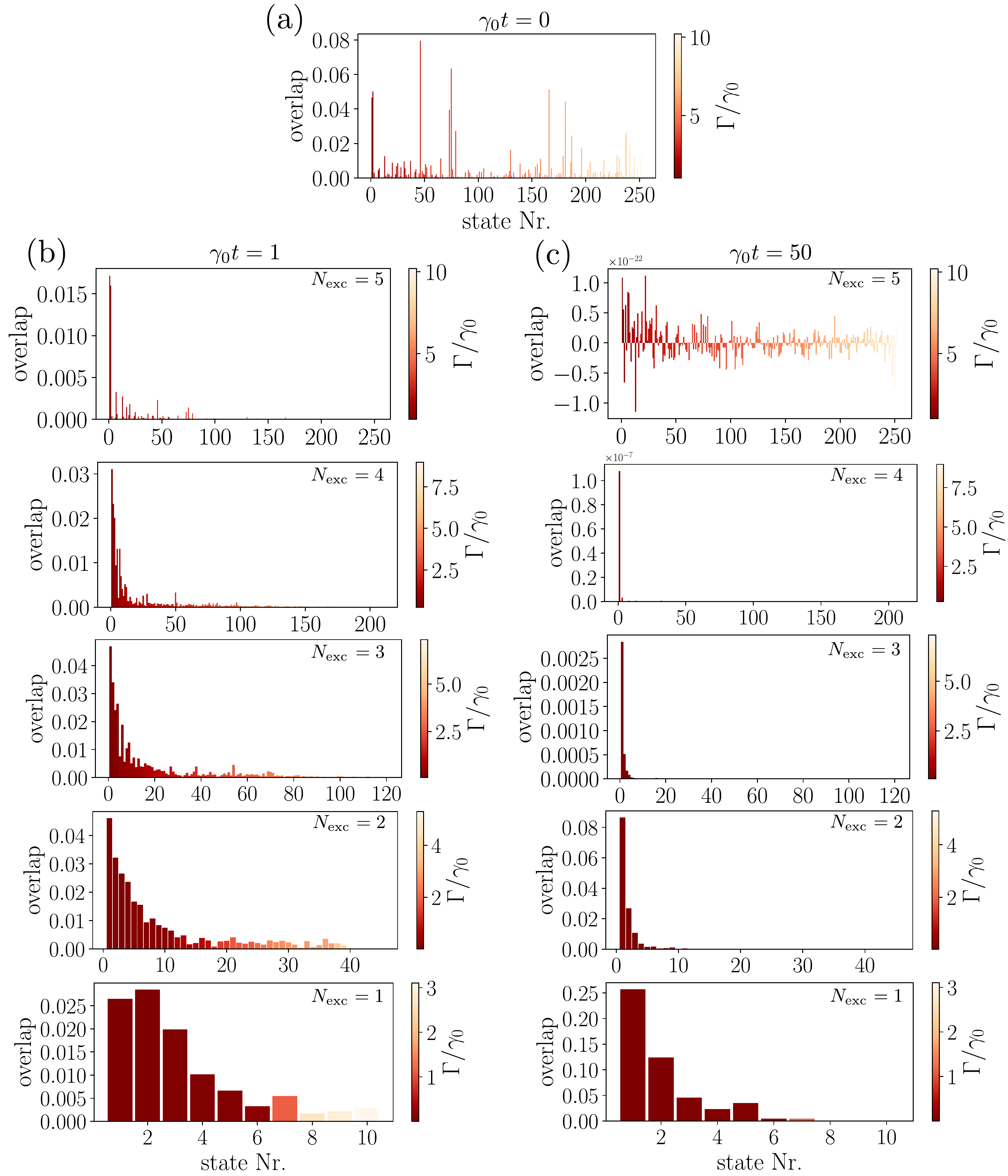}
    \caption{Overlaps of the dynamically generated state $\hat{\rho}(t)$ at different times with all individual states $\ket{\psi_i}$ constituting an excitation manifold $\Psi_{N_\mathrm{exc}}$ for a ten-atom chain ($a = 0.15 \lambda_0$) with half the atoms initially excited in a checkerboard pattern. (a) Overlaps with all states constituting the five-excitation manifold at $\gamma_0 t=0.0$ (the overlaps with all other manifolds are zero at the initial time). Panels (b) and (c) show the overlaps with all states for $N_\mathrm{exc}\in \{1...5\}$ at (b) $\gamma_0 t = 1$ and (c )$\gamma_0 t =50$. In each panel, the states are sorted from left to right with increasing collective decay rate $\Gamma$, whose exact value is indicated by the color code. All parameters are the same as in Fig.\,2.}
    \label{fig:incoh_ind_state_overlaps}
\end{figure}
In~\fref{fig:incoh_ind_state_overlaps}, we present the overlaps with the individual states making up each excitation manifold for the incoherent checkerboard excitation $|\psi_\mathrm{incoh} \rangle$ presented in Fig.\,2(a) at different times. The initial state exhibits overlaps with many states contained in the $N_\mathrm{exc}=5$ manifold. In contrast to the case of a coherent excitation, however, the dynamically generated state predominantly exhibits non-zero overlaps with the least radiative states contained in the multi-excitation manifolds already at early times [$t \gamma_0 = 1$; see~\fref{fig:incoh_ind_state_overlaps}(b)]. This fundamental difference in the state's dynamics, which becomes more pronounced at later times [see~\fref{fig:incoh_ind_state_overlaps}(c)], is key to the transient generation of multi-excitation subradiant states presented in this work.

\section{Extracting late time effective decay rates}
From Fig.\,2(a) in the main text, it is evident that the excited state population in each manifold decays relatively fast despite having predominantly populated the subradiant states in each manifold [see overlap data in Fig. 2(b)]. In this section, we discuss that this effect is not a shortcoming of the employed preparation mechanism, but an intrinsic feature of dipole-coupled systems. Diagonalizing the non-Hermitian Hamiltonian of the system for the different excitation manifolds, we find that the most subradiant states in higher excitation manifolds have larger decay rates than the single excitation subradiant states. This naturally results in a faster decay of the population trapped in higher excitation manifolds.

\begin{figure}
    \centering
    \includegraphics[width = 0.8\textwidth]{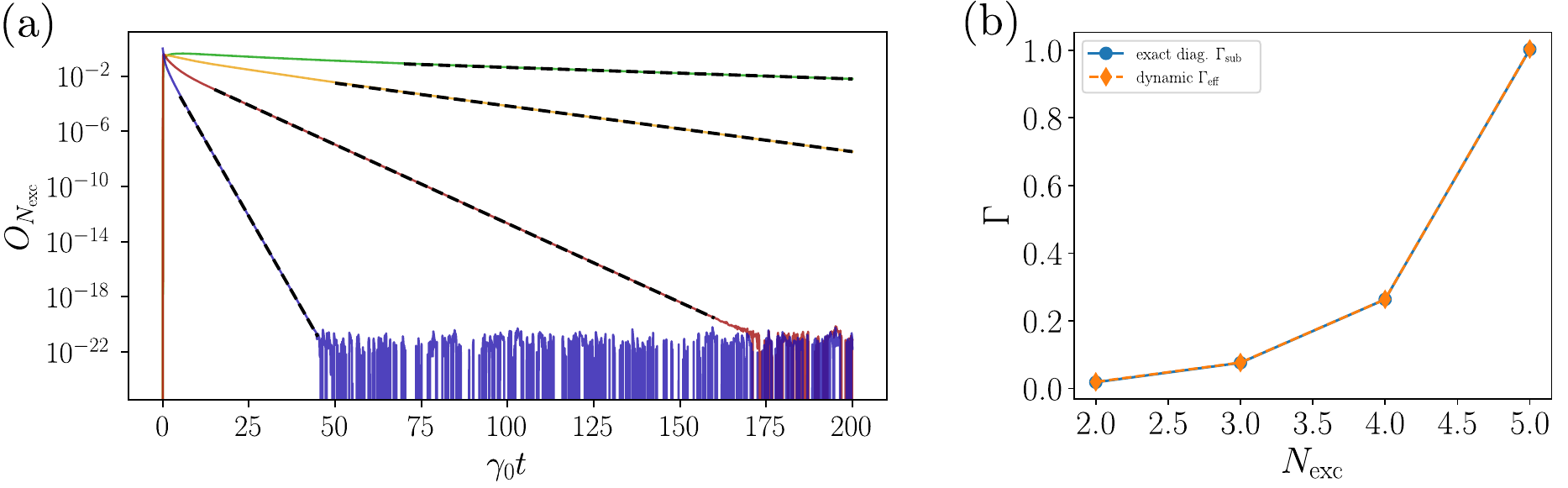}
    \caption{(a) Overlap with the different excitation manifolds over time for a ten-atom chain ($a = 0.15 \lambda_0$), with half the atoms initially excited in a checkerboard pattern. The effective decay rates of the exponential decay can be extracted from linear fits of the population in logarithmic scale. The different colors represent the different excitation manifolds: $N_\mathrm{exc} = 5$ (blue), $N_\mathrm{exc} = 4$ (red), $N_\mathrm{exc} = 3$ (yellow) and $N_\mathrm{exc} = 2$ (green). (b) Comparison of the effective decay rates $\Gamma_\mathrm{eff}$ obtained in panel (a) with the decay rate $\Gamma_\mathrm{sub}$ of the most subradiant state contained in the each excitaion subspace.}
    \label{fig:gamma_comparison}
\end{figure}

This increasing decay rate for larger excitation manifolds is nicely observed when plotting the populations in logarithmic scale [see Fig.~\ref{fig:gamma_comparison}(a)]. Performing linear fits to the late time dynamics of each manifold, we can extract the effective decay rate of the corresponding exponential decay. As shown in Fig.~\ref{fig:gamma_comparison}(b), they coincide with the decay rate of the darkest eigenstate in each manifold. This demonstrates that the smallest decay rates of any specific configuration of interacting emitters can be reached with the presented protocol.

\section{Incoherent checkerboard state preparation}
One of the fundamental results of our work is that diminishing coherences between the atoms at initial times results in a more efficient transient generation of many-body subradiant states. Here, we show how the optimal incoherent checkerboard state $|\psi_\mathrm{CB} \rangle = \prod_{i=n}^N \hat{\sigma}_{2i+1}^{eg} |G \rangle$ can be prepared for a six atom chain by applying a global \emph{coherent} laser drive and a local detuning pattern.

To model this state preparation scheme, we introduce two additional terms in the Hamiltonian given by Eq.\,(2), which describe the global laser drive at a Rabi frequency $\Omega$ and the local position-dependent detunings $\Delta_n$ from the bare atomic transition frequency
\begin{equation}
    \hat{\mathcal{H}}_{\Omega} = \hbar \sum_{n=1}^N (\omega_0 + \Delta_n) \hat{\sigma}_n^{ee} + \hbar \Omega \sum_{n=1}^N (\hat{\sigma}_n^{eg} + \hat{\sigma}_n^{ge}) + \hbar \sum_{n,m \neq n}^N J_{nm}  \hat{\sigma}_n^{eg} \hat{\sigma}_m^{ge}.
    \label{eqn:drive_ham}
\end{equation}
Setting $\Delta_{2n+1} \equiv \Delta$ and $\Delta_{2n} = 0$ imposes a checkerboard detuning pattern onto the atoms. Such a pattern can for example be generated by trapping the atoms in a superlattice instead of a plain optical lattice. Local AC Stark shifts then result in the required detuning pattern.

\begin{figure}
    \centering
    \includegraphics[width = 0.8\textwidth]{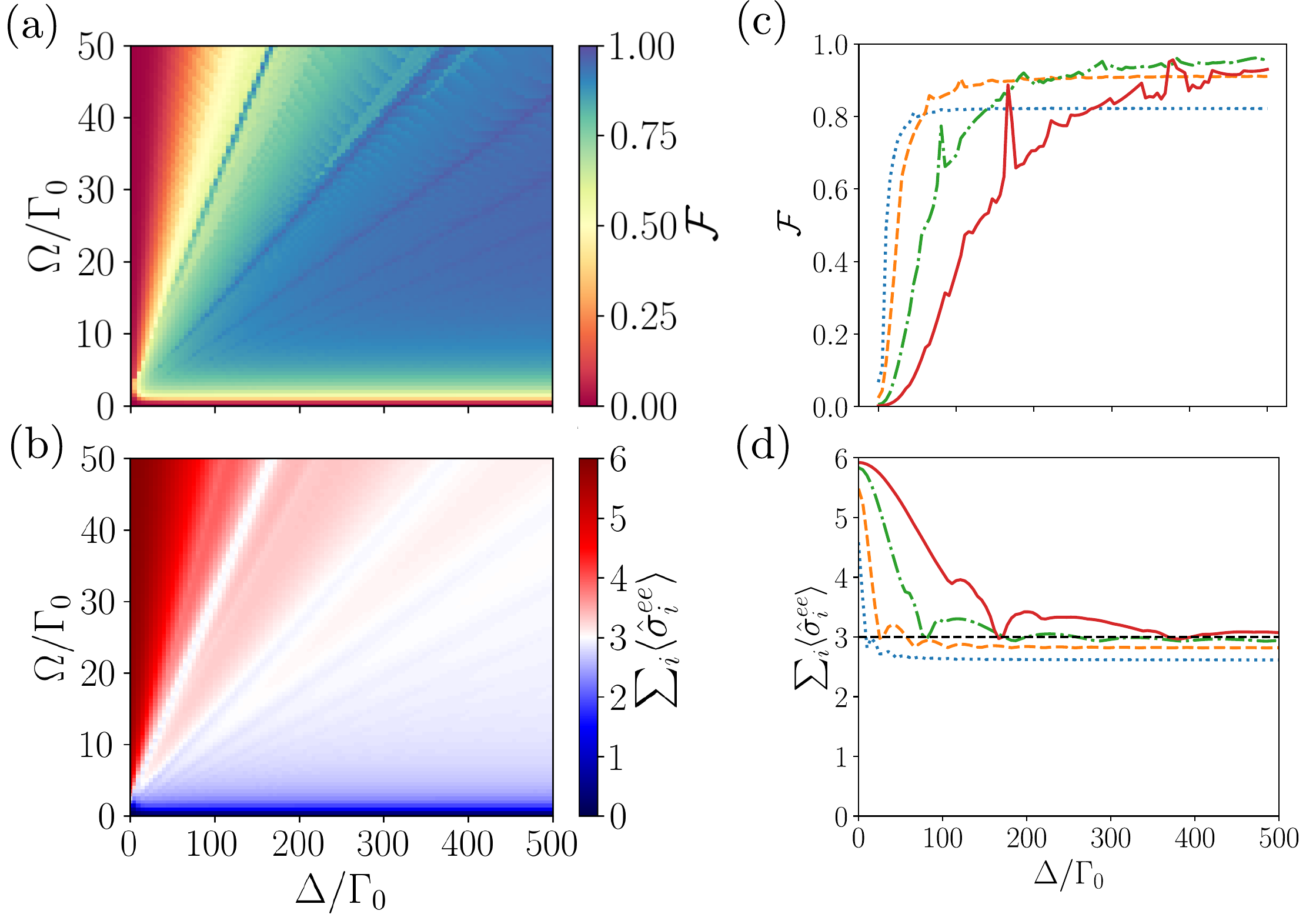}
    \caption{Dynamic preparation of the incoherent checkerboard state $\rho_\mathrm{CB}$ for a six atom chain. (a) Fidelity $\mathcal{F} \equiv \text{tr}\left(\sqrt{\sqrt{\rho_\mathrm{CB}} \rho_\mathrm{dyn} \sqrt{\rho_\mathrm{CB}}}\right)$ and (b) total excited state population in the atomic ensemble as a function of Rabi frequency $\Omega$ and detuning $\Delta$. (c) Fidelity and (d) total excited state population as a function of detuning $\Delta$ for $\Omega/\Gamma_0 = 5$ (blue dotted line), $\Omega/\Gamma_0 = 10$ (orange dashed line), $\Omega/\Gamma_0 = 25$ (dash dotted green line) and $\Omega/\Gamma_0 = 50$ (solid red line). }
    \label{fig:state_prep}
\end{figure}

We simulate the time evolution of the master equation given in Eq.\,(1) of the main text for the Hamiltonian~\eqref{eqn:drive_ham} and for different values of $\Delta$ and $\Omega$. This allows us to determine the state $\rho_\mathrm{dyn}$ at the time at which the ensemble is maximally inverted. We then compare this dynamically generated state with the desired initial condition $\rho_\mathrm{CB} \equiv \ket{\psi_\mathrm{CB}}\bra{\psi_\mathrm{CB}}$ by calculating the fidelity $\mathcal{F} \equiv \text{tr}\left(\sqrt{\sqrt{\rho_\mathrm{CB}} \rho_\mathrm{dyn} \sqrt{\rho_\mathrm{CB}}}\right)$, and determine the parameter regime for $\Delta$ and $\Omega$ for which the incoherent checkerboard state can be prepared efficiently. The results are shown in~\fref{fig:state_prep}.
We find that there exists a large parameter regime where the checkerboard state can be prepared with fidelities $\mathcal{F}>90\%$. Intuitively, this region corresponds to large drive compared to the decay rate such that the first Rabi oscillation can reach close to full inversion. Additionally, the detuning needs to be large compared to the drive to avoid the undesired population of the detuned emitters. It can be numerically shown that such an inexact preparation of the initial incoherent checkerboard state still results in efficient transient generation of the subradiant states.

Finally, it is worth noting that this preparation scheme does not require local atom addressing. It simply requires to apply a global drive with an addtional local detuning pattern, which can be achieved in state-of-the-art experiments via optical superlattices \cite{Superlattices_1,Superlattices_2}. In principle, this procedure can be extended to generate arbitrary configurations of initially excited atoms. Our work therefore outlines anexperimentally viable path towards the transient generation of multi-excitation subradiant states.

\section{Coherent dynamics: atomic detuning pattern}
\begin{figure}
    \centering
    \includegraphics[width=0.7\textwidth]{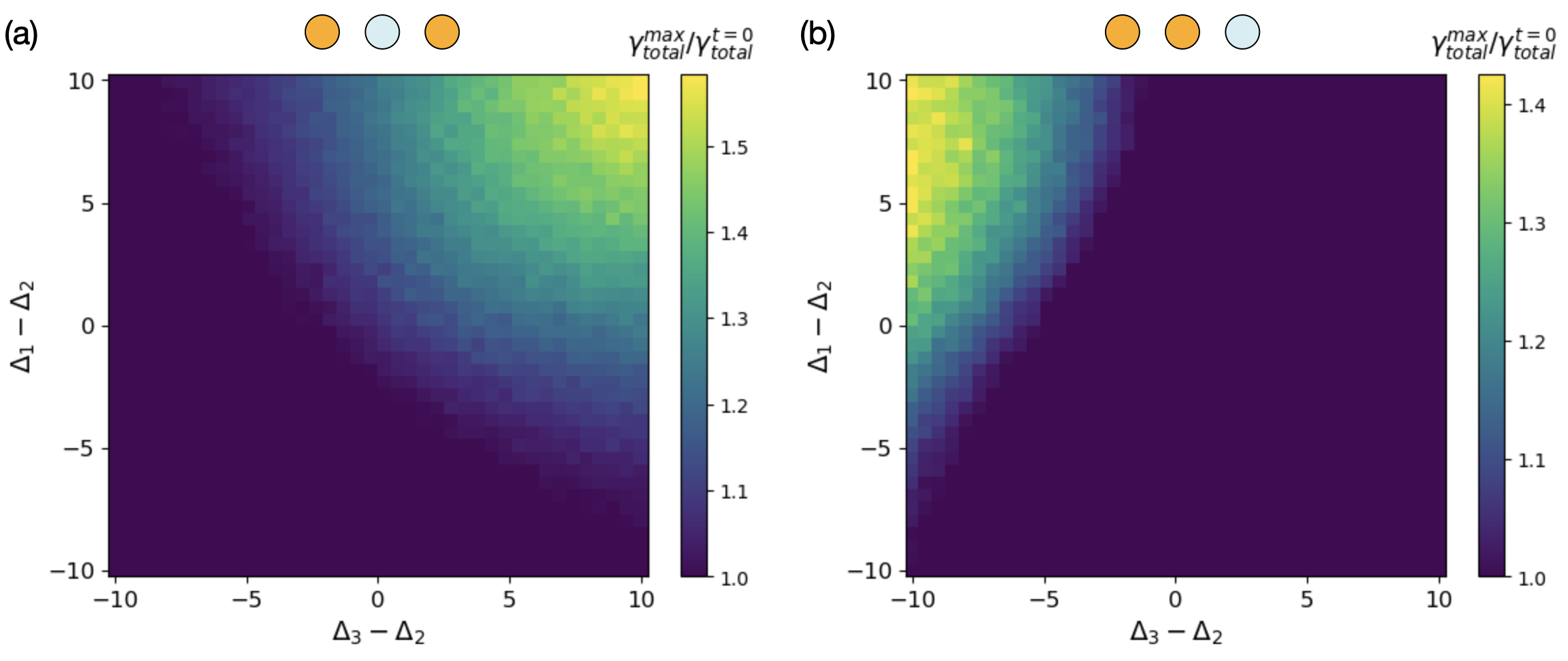}
    \caption{Maximum emission rate divided by initial emission rate, $\gamma_ \mathrm{total}^ \mathrm{max}/\gamma_ \mathrm{total}^{t=0}$, as a function of the relative detuning between three atoms in a chain, $\Delta_1-\Delta_2$ and $\Delta_3-\Delta_2$. In (a), the two atoms at the edges are initially excited (drawn in orange), whereas two neighboring atoms are initially excited in (b). In both cases, a spacing of $a=0.075\lambda_0$ is considered.}
    \label{fig:appendix_detuning}
\end{figure}

In the main text, we show that coherent dipole-dipole interactions generally couple the states within the same excitation manifold. For strong enough interactions ---that is, small enough lattice spacings---, this can give rise to a significant population transfer from subradiant to superradiant states, which can ultimately result in a radiation burst driven by the coherent dynamics of the system ---as opposed to the typical Dicke superradiance, which purely emerges from the dissipative interactions in the system---. Here, we demonstrate that similar effects appear when modifying the resonance frequency of different atoms. For that, we again consider a more general Hamiltonian than that in Eq.\,(2),
\begin{equation}
\label{eq:Hamiltonian_local_detunings}
    \hat{\mathcal{H}}_{\Delta} = \hbar \sum_{n=1}^N (\omega_0 + \Delta_n) \hat{\sigma}_n^{ee} + \hbar \sum_{n,m \neq n}^N J_{nm}  \hat{\sigma}_n^{eg} \hat{\sigma}_m^{ge}, 
\end{equation}
which includes position-dependent detunings $\Delta_i$ from the bare transition frequency of each atom. For a three-atom chain, for example, the dynamics now additionally depend on the frequency differences $\Delta_1-\Delta_2$ and $\Delta_3-\Delta_2$. In~\fref{fig:appendix_detuning}, we plot the maximum emission rate divided by the initial emission rate, $\gamma_ \mathrm{total}^\mathrm{max}/\gamma_ \mathrm{total}^{t=0}$, as a function of the detuning differences for arrays with spacing $a=0.075\lambda_0$. If the first and third atoms are initially excited [\fref{fig:appendix_detuning}(a)], a radiation burst is already present at zero detuning ($\gamma_\mathrm{tot}^\mathrm{max} \approx 1.07 \gamma_\mathrm{tot}^{t=0}$). For $\Delta_1-\Delta_2>0$ and $\Delta_3-\Delta_2>0$, the detuning pattern induces a population transfer from subradiant states to radiant states at initial times and the magnitude of the radiation peak consequently increases. For other combinations of the atomic detunings, this transfer is suppressed and the peak vanishes.

The specific choice of detunings depends on the configuration of the ensemble. A system with the first and second atoms initially excited, for instance, presents a radiation burst for $\Delta_1-\Delta_2>0$ and $\Delta_3-\Delta_2<0$, as shown in~\fref{fig:appendix_detuning}(b). In this case, additionally, no peak is observed at zero detuning.

Note that the required detuning pattern can be readily realized in state-of-the-art experiments. In experiments involving cold atoms, this can be achieved by superimposing multiple optical lattices with varying periodicities, that is, by creating optical superlattices \cite{Superlattices_1,Superlattices_2}.
For optical tweezers, intensity modulations of the individual tweezer beams can be employed to generate alternating AC-Stark shifts on the atoms.

\end{document}